\documentstyle[preprint,aps,epsf]{revtex}
\begin{document}
\draft
\title{ Universal Pulse Shape Scaling Function and Exponents: A Critical test for
	Avalanche Models applied to Barkhausen Noise }
\author{ Amit P. Mehta, Andrea C. Mills, Karin Dahmen, James P. Sethna$^\dag$}
\address{ Department of Physics, University of Illinois at Urbana-Champaign, Urbana, IL 61801 }
\address{ $^\dag$Laboratory of Atomic and Solid State Physics (LASSP), Clark Hall, 
	Cornell University, Ithaca, NY 14853 }
\date{\today}
\maketitle
\begin{abstract}
	In order to test if the universal aspects of Barkhausen noise in magnetic materials can be
	predicted from recent variants of the non-equilibrium zero temperature Random Field Ising Model (RFIM),
	we perform a quantitative study of the universal scaling function derived from the 
	Barkhausen pulse shape in simulations and experiment.  Through data collapses and scaling relations
	we determine the critical exponents $\tau$ and $1/\sigma\nu z$ in both simulation
	and experiment.  Although we find agreement in the critical exponents, we find differences
	between theoretical and experimental pulse shape scaling functions as well as
	between different experiments.
\end{abstract}


\section{ Introduction } 

	Real materials, have dirt, or disorder, which often leads to slow (glassy) dynamics due to
	complex free energy landscapes with diverging energy barriers\cite{barrier}.  
	On long length scales and practical time scales, thermal effects often
	become unimportant. Indeed, when such a system is driven by an external field it
	jumps from one local metastable free energy minimum to the next, and the state
	of the system depends on its history.  

	In magnetic materials jumps from one local minimum to the next involve a collective process whereby
	clusters of magnetic domains change the direction of their magnetization, in an avalanche.  
	These avalanches can be triggered by a slowly but continuously increasing homogeneous
	external magnetic field $H$ (taken from $-\infty$ to $+\infty$). 
	These avalanches produce so-called Barkhausen noise, which can be observed experimentally as a voltage 
	signal induced in a pickup coil wound around the magnet. Experiments show that these avalanches
	come in all sizes; their sizes are typically distributed according to a power law over several decades.  
	Other systems also exhibit a broad range of avalanche sizes and durations following power laws:
	superconducting vortex line avalanches\cite{Field}, resistance avalanches in a superconducting 
	film\cite{Wu}, capillary condensation of helium in Nuclepore\cite{Nuclepore}, acoustic 
	emissions in athermal martensites\cite{Ortin}, earthquakes\cite{GR}, and many
	others\cite{Nature}.	

	We examine two variants of the non-equilibrium
	zero temperature random field Ising model (RFIM), which both predict
	power law distributions of avalanche sizes and {\em universal} 
	non-equilibrium collective behavior.  The RFIM is a model for a conventional magnet
	in which magnetic domains are modeled by ``spins'' on a lattice that can only point
	up or down.  The first variant basically exhibits single front
	(domain wall) propagation
	dynamics in which spins at the edge of an existing front flip when it is (locally) energetically
	favorable to do so.  Spins that are not adjacent to this front are very unlikely to
	 flip on their own due to the presence of an infinite-range demagnetizing field, which
	is present in addition to nearest neighbor ferromagnetic interactions. We call this the front
	propagation model\cite{Matt1}.
	In the second variant, called the nucleation model, spins flip anywhere in the system
	when it is (locally) energetically favorable to do so. In this case there 
	are {\em many interacting} fronts, unlike in the front propagation model. Only nearest 
	neighbor ferromagnet interactions are included in the nucleation model\cite{JPSethna}.

	In this paper we perform a detailed quantitative comparison of the universal avalanche pulse 
	shapes and exponents
	obtained from: front propagation dynamics, domain nucleation dynamics, and experiment. Our
	analysis constitutes a test of whether the non-equilibrium zero-temperature RFIM (either
	variant) is in the same universality class as experimental systems exhibiting Barkhausen noise.

\section{ The Model } 

	The RFIM consists of a (hypercubic) lattice of $N$ spins
	($s_i = \pm 1$), which may point up ($s_i = +1$) or down ($s_i = -1$). Spins are coupled to 
	nearest neighbors (through a ferromagnetic
	exchange interaction $J$), and to an external field $H(t)$ which is increased
	adiabatically slowly.  To model dirt in the material, we assign a random field, $h_i$, to
	each spin, chosen from a distribution $P(h_i) = exp(-h^2_i/2R^2)/\sqrt{2\pi}R$,
	where $R$, the disorder, determines the width of the Gaussian probability distribution and
	therefore gives a measure of the amount of quenched disorder for the system. 

	The
	Hamiltonian for the system at a time $t$ is given by:
	$H = \sum\limits_{<ij>} -J s_i s_j - \sum_{i}(H(t) + h_i - J_{inf}M)s_i$,
	where $J_{inf}$ is the strength of the infinite range demagnetizing field 
	($J_{inf} = 0$ for the nucleation model), 
	$M = \frac{1}{N}\sum\limits_{i}s_i$ is the magnetization of the system, 
	and $<ij>$ stands for nearest neighbor
	pairs of spins. Initially, $H(-\infty) = -\infty$ and
	all the spins are pointing down. Each spin is always aligned with its local effective 
	field $h_i^{eff} = J\sum_{<ij>}s_j + H(t) + h_{i} - J_{inf}M$.  The external field
	$H(t)$ is adiabatically slowly increased from $-\infty$ until the local field, 
	$h_i^{eff}$, of any spin $s_i$ changes sign, causing the spin to flip\cite{Matt1,Sethna}.  
	It takes some microscopic time $\Delta t$ for a spin to flip 
	($\Delta t \equiv 1$ for our simulation). The spin flip changes the local field of the 
	nearest neighbors and may cause them to flip as well, etc.  This {\em avalanche} process
	continues until no more spin flips are triggered.
	Each step of the avalanche, that is, each $\Delta t$ in which a set of spins simultaneously flip,
	is called a {\em shell}. The number of spins that flip in a shell is directly proportional to
	the voltage $V(t)$ during the interval $\Delta t$ that an experimentalist would measure in a
	pick-up coil wound around the sample. In our simulations we therefore denote the number
	 of spins flipped in a shell at a time $t$ by $V(t)$.   The first shell of an avalanche
	 (one spin flip) is triggered by the external field
	$H(t)$, while each subsequent shell within the avalanche is triggered only by the previous shell, since 
	$H(t)$ is kept constant while the avalanche is propagating.  $H(t)$ is only increased when the 
	current avalanche has stopped, and is increased only until the next avalanche is triggered
	(i.e. $\frac{dH}{dt}\rightarrow 0$). 
	The number of shells in an avalanche times
	$\Delta t$ defines the {\em pulse duration}, $T$, or the time it took for the entire avalanche to flip.
	In this paper we will be interested in looking at $V(t,T)$ for $0 < t \le T$, that is,
	the voltage as a function of time for an avalanche of a given duration $T$.	

	The front propagation model exhibits self-organized criticality (SOC)\cite{Urbach,Narayan,Zapperi}.  
	This means that as $H$ is increased the model always operates at the critical depinning point, 
	and no parameters need to be tuned to exhibit critical scaling behavior (except
	$\frac{dH}{dt}\rightarrow 0$).
	The nucleation model, on the other hand, is a plain old critical system with a continuous 
	second order phase transition with disorder as the tuning parameter. 
	The continuous non-equilibrium phase transition can be
	 understood as follows. For zero disorder, the random fields of each spin will 
	be the same, so that when one spin is flipped the entire lattice
	of spins will flip. This results in a rectangular hysteresis curve with a macroscopic jump
	in the magnetization when the external field overcomes the interaction with the
	neighbors. On the other hand, if the disorder is infinite, each spin will have
	 a very different random field, so as 
	the external field is raised each spin will essentially flip independently, triggering 
	no other spins to flip.  This will result in a smooth ``hysteresis'' curve with an 
	approximately constant slope ($M(H)\sim H$ over a wide range of $H$).
	In between these two phases of behavior there is a continuous phase transition at some 
	critical value $R = R_c$. At the transition each branch of the hysteresis loop
	has a single point with infinite
	slope ($\frac{dM}{dH}\mid_{\pm H_c}\rightarrow\infty$) at a critical field 
	$H = \pm H_c$. $R_c$ and $H_c$ are nonuniversal. In three dimensions they are: $R_c = 2.16$ and
	$H_c = 1.43$ (in units of $J$)\cite{Olga1}. The transition is characterized by a number
	of universal critical exponents \cite{Olga2}, and scaling functions.
	In this paper we specifically focus on only two exponents: $1/\sigma\nu z$ and $\tau$,
	and one scaling function. The significance of these exponents is given later. 
	The scaling function we examine can be obtained through experiment as well as simulation.
	Details of the simulation algorithm are given elsewhere\cite{Matt2}. 			

\section{ The Experiments }

	In addition to examining results obtained from simulation we 	
	study results from three different experiments. We performed
	one of these experiments, and the results for the other two experiments 
	were obtained from already published results\cite{Spaso,Durin}.

	Our own experiment was performed on an (unstressed) amorphous alloy,
	$Fe_{21}Co_{64}B_{15}$.  The data we present from Durin et al\cite{Durin} is 
	from an experiment on an amorphous alloy with
	a different composition, $Fe_{64}Co_{21}B_{15}$, under a tensile stress.
	Studies of the effect of tensile stress on samples of this type indicate
	that for low tensile stress, the domain structure is a complicated pattern
	of maze domains, dominated by quenched-in stresses\cite{Durin1}.  
	On the other hand, when such materials
	undergo tensile stress, the uniaxial anisotropy gives way to a simpler
	domain structure with a few parallel domains in the direction of the
	stress\cite{Durin1}.  Related to this
	change in domain structure is a change in the dominant interaction in
	the material.  In amorphous alloys under stress, surface tension effects
	are thought to be more important than dipolar interactions, while dipolar
	interactions dominate for polycrystals and materials with small grains
	\cite{Durin2}. 

	Durin et al's sample was 
	under stress so as to enhance stress-induced anisotropy so much that 
	the long range dipolar interactions can be neglected, placing their experiment into the 
	universality class of the front propagation model\cite{Durin2,Durin3,Robbins}.
	Spasojevi\'{c} et al's sample was a quasi-two-dimensional metal glass, more precisely a 
	commercial VITROVAC 6025 X\cite{Spaso}.  Based on the scaling exponents obtained from
	Spasojevi\'{c} et al's experiment (given later), their experiment does not seem to 
	fall into any universality class discussed in this paper.
	Our experiment seems to be in a crossover regime between two universality
	classes; details are given later in this paper.
	Further details about Durin et al's, and 
	Spasojevi\'{c} et al's experiment can be obtained elsewhere
	\cite{Spaso,Durin,Durin1,Durin2}.

\section{ Extracting Pulse Shapes } 

	{\em Nucleation Model}: We simulate four realizations of a $1200^3$ system near $R = R_c$ 
	($R = 2.2$) and record 
	the time series $V(t,T)$ of avalanches from an $H$ window near $H_c$ ($1.42 < H < 1.43$).   
	We average avalanches of a fixed pulse duration $T$ (within the interval $[T,1.05T]$), 
	for various values of $T$.  In each case we average over $1000$ to 
	$2000$ avalanches to ensure strong fluctuations have been averaged out.  We check finite size effects
	by performing simulations at $800^3$ and $1000^3$ and verifying that identical avalanche shapes
	(within small fluctuations) are obtained for all system sizes.  

	{\em Front Propagation Model}: We perform $100$ realizations of a $400^3$ system and record avalanches
	from an $H$ window within the slated part of the hysteresis loop ($1.25 < H < 1.88$).  Even though 
	the front propagation model exhibits SOC due to the infinite range demagnetizing interaction, 
	in order to avoid effects due to initial nucleation of the front (beginning of the
	hysteresis loop), or when the front encounters the boundaries of the simulation (the end
	of the hysteresis loop) we must choose avalanches near the middle of the hysteresis loop. 
	We obtain avalanche shapes $V(t,T)$ in a manner identical to how they were
	obtained for the nucleation model. We check finite size effects by performing simulations of $200^3$
	and $300^3$ size systems, and find a consistent avalanche shape for all three system sizes. 

	{\em Experiment}: 
	Measurements were performed on a 21 cm x 1 cm x 30 $\mu$m ribbon of
	$Fe_{21}Co_{64}B_{15}$ alloy, a soft amorphous ferromagnet obtained from
	Gianfranco Durin.  The domain walls run parallel to the long axis of the
	material, with about 50 domains across the width.
	A solenoid, driven with a triangle wave, applies a magnetic field along
	the long axis of the sample.  Since domain wall motion dominates over
	other means of magnetization in the linear region of the loop, data were
	collected in only a selected range of applied fields near the center of
	the loop.
	The Barkhausen noise was measured by a small pick-up coil wound around the
	center of the sample.  This voltage signal was amplified, anti-alias
	filtered and digitized, with care taken to avoid pick-up from ambient
	fields.
	Barkhausen noise was collected for both increasing and decreasing fields
	for $80$ cycles of the applied field through a saturation hysteresis loop.
	The driving frequency was $0.01$ Hz; this corresponds to $c=0.09$, where c is
	a dimensionless parameter proportional to the applied field rate and is
	defined in the Alessandro Beatrice Bertotti Montorsi model (ABBM model) for 
	the Barkhausen effect\cite{ABBM}. In this way, our measurements should be
	well inside the $c<1$ regime identified in the ABBM model, in which we 
	can expect to find separable avalanches rather than continuous domain 
	wall motion.

	Due to background noise in real experimental data there is no definitive way to
	determine when an avalanche begins and when one ends; we set a sensitivity threshold which is
	high enough to cut out background noise and low enough to capture 
	enough of the avalanche so as not to affect the shape.  We check
	the validity of our threshold by perturbing the threshold by a small 
	amount and noticing that there is no change in the avalanche shape.  

\section{ Critical Exponents and Data Collapses }

	In the front propagation model, as well as in the nucleation model near ($R_c,H_c$), the 
	voltage $V(T,t)$ scales as\cite{Matt1}:

\begin{equation} V(T,t) = T^{1/\sigma\nu z-1}f_{shape}(t/T) \end{equation}

	By collapsing average avalanche shapes of various durations $T$ we determine the universal 
	scaling function, $f_{shape}(t/T)$, and the critical exponent $1/\sigma\nu z$.
	The exponent $1/\sigma\nu z$ relates the avalanche size, $S$, to the avalanche
	pulse duration, $T$, at criticality by $S \sim T^{1/\sigma\nu z}$. 
 	We find that the critical exponent $1/\sigma\nu z$ for the front propagation
	and nucleation model, obtained from simulation, is in close agreement 
	with previous theoretical predictions and with experimental values\cite{Dahmen}. 
	The collapses are shown in Figs. 1 and 2. 

	We also determine the avalanche 
	size distribution that scales as $D(S) \sim S^{-\tau}$ at criticality.  The
	avalanche size distributions for our simulation and our experiment are given in
	Fig. 3; the values obtained for $\tau$, for the front propagation model and
	nucleation model, are in close agreement with previously quoted values\cite{Matt1,Olga1}.
	Experimentally, the scaling exponent $\tau$ weakly depends on $c$ in some materials
	\cite{Durin1}. From this dependence we find that there may be a difference 
	of $0.02$ between the value of $\tau$ we find in our experiment and the
	value of $\tau$ at $c = 0$ (zero frequency).  However, this difference is
	within the error bars we give for $\tau$. Table I summarizes results for 
	$\tau$ and $1/\sigma\nu z$ for experiment, simulation, and mean field theory.  
	We see that the critical exponents for the front propagation model agree 
	(within error bars) with exponents from Durin et al's experiment, as expected.

	While Durin et al's experiment is believed to fall into the front propagation
	universality class, as discussed above, our experiment is neither in the 
	front propagation nor in the mean field universality class.  A priori
	we would assume that our experiment would be in the mean field universality 
	class since our sample was unstressed, and previous experiments have indicated
	that unstressed samples will exhibit mean field behavior\cite{Durin2}.
	The critical exponents found from our experiment indicate that it maybe in a crossover
	regime between the mean field and the front propagation universality classes. The 
	exponent $\tau$ has a value of $1.46\pm0.05$ (see also Table I), which is between 
	the value of $1.28$ for the front propagation model 
	(sample with stress) and the mean field value of $1.5$ (sample without stress)\cite{Matt1,Durin2}.  
	In addition, the exponent $\alpha = 1.74\pm0.06$, determined from the
	avalanche pulse duration distribution given by $D(T) \sim T^{-\alpha}$ at criticality,
	(see Inset of Fig. 2), is between the value of $1.5$ for the front propagation 
	model (with stress), and the mean field value of $2$ (without stress)\cite{Durin}. 
	Residual stress on our 
	sample may have resulted in these anomalous exponents.

	Although our experiment may be in a crossover regime, we were able to obtain a good
	collapse of the avalanche pulse shapes (see Fig. 2). In order to reaffirm
	the validity of the exponent obtained from the collapse, we independently
	checked the value of $1/\sigma \nu z$ from the power spectra 
	($P(w) \sim w^{-1/\sigma \nu z}$ at criticality), and found 
	$1/\sigma \nu z = 1.73 \pm 0.08$, which is consistent (within errors bars)
	with $1/\sigma \nu z = 1.70 \pm 0.05$ obtained from the 
	experimental avalanche pulse collapse.

\section{ Fitting to Orthonormal Polynomials }

	The mean field shape of avalanches in our model $V^{MFT}(t,T)$ is an inverted parabola
	\cite{KSprivate}. In order to study corrections to the mean field shape, 
	in simulation and experiment, 
	we derive a set of orthonormal polynomials {$f_i(t)$} with $f_i(-L) = f_i(L) = 0$, 
	where $L = T/2$ is half the duration of the avalanche. 
	The negative of the first polynomial, $f_0(t)$, is proportional to the mean field 
	result. The first four polynomials of the set are:

\begin{eqnarray}
	&& f_0(t) = \sqrt{\frac{15}{16L^5}}(t^2 - L^2)\\
	&& f_1(t) = \sqrt{\frac{105}{16L^7}}(t^3 - L^2t)\\
	&& f_2(t) = \sqrt{\frac{45}{64L^9}}(7t^4-8L^2t^2+L^4)\\
	&& f_3(t) = \sqrt{\frac{1155}{64L^{11}}}(3t^5-4L^2t^3+L^4t)\\
	&& f_4(t) = \sqrt{\frac{1365}{2048L^{13}}}(33t^6-51L^2t^4+19L^4t^2-L^6)
\end{eqnarray}

	We fit a linear combination
	of the above polynomials to the average avalanche shape obtained from simulation and
	experiment. $L$ is also left as a free parameter in the fits to the average shapes. 
	The pulse duration is then precisely defined as $T = 2L$. 
	The results of the fits are shown in Fig. 1 and Fig. 2.  In Fig. 4 we give the
	coefficients for the fits, found in simulation and experiment.  We also include the
	coefficients for the fit to an average avalanche shape determined by Durin et al's experiment 
	\cite{Durin}, whose avalanche shape is given in Fig. 5. The coefficients are
	determined from the total fit function:
	
\begin{equation} F(t) = a_0f_0(t)+a_1f_1(t)+a_2f_2(t)+a_3f_3(t)+ a_4f_4(t) \end{equation}
	
	$a_0$, $a_2$, and $a_4$ are the {\em symmetric } coefficients of the fit, while
	$a_1$, $a_3$ are the {\em antisymmetric } coefficients of the fit (i.e. multiplying polynomials
	that are not symmetric under time reversal $t\rightarrow-t$).  We are 
	particularly interested in the asymmetry of the avalanche shapes.

	From inspection of the avalanche shapes we see that the experimental avalanche
	shapes are strongly asymmetric under time reversal, while the avalanche shapes
	determined from the simulation of the two models
	are both very close to symmetric.  Quantitatively, the coefficients for the
	avalanche shapes of the nucleation model are very similar to those of the
	front propagation model and both are different from the  
	experimental avalanche shapes' coefficients. While avalanche shapes in both models are 
	slightly asymmetric to the left (i.e. $V(t,T)$ increases more slowly than it decreases), 
	the experimental avalanche shapes are strongly asymmetric to the right direction 
	(i.e. $V(t,T)$ increases quickly and decreases slowly). The origin of this difference
	between theory and experiment is not yet understood. It comes as a surprise, since the 
	critical exponents obtained in Durin et al's experiment are in close agreement with exponents
	obtained from the front propagation model (see Table I) and the avalanche shapes are expected 
	to be just as universal as the critical exponents.  On the other hand, we see in Fig. 5 that 
	different experiments give different pulse shape scaling functions.  This difference may
	be a result of the fact that the experiments are not in the same universality class. 
	Nevertheless, all the experimental pulse shapes do have the same sign of asymmetry.

	We did one check on whether the pulse shape function in our model is indeed {\em universal}, 
	an assumption often taken for granted: We changed the lattice of our simulation from a simple cubic 
	to a BCC (body centered cubic) lattice and found an identical pulse shape scaling function
	(within small statistical error), as expected when universality holds.

\section{ discussion and concluding remarks }

	In performing these analyses we have stumbled upon an interesting observation:
	after appropriate rescaling of the y-axis, the universal pulse shape function for
	the front propagation and nucleation model appear to 
	look the same (see Inset for Fig. 1) even though
	we know \cite{Nature} that the models do not belong to the same universality class.  However,
	upon more precise quantitative analysis we see that they are in fact different, their
	$a_1$ coefficients are different by several $\sigma$.  Furthermore, we tried to collapse two
	pulse shape functions with $T \simeq 52$ from the two different models and found that they
	could not be collapsed precisely despite, naturally, scaling by different critical exponents 
	appropriate for the two models. The front propagation pulse scaling function
	is more asymmetric, as supported by the somewhat larger $a_1$ fit coefficient compared
	to the nucleation model result.

	By examining the pulse shape scaling functions as a sharper test than merely
	critical exponents for the universality class of the non-equilibrium zero
	temperature RFIM, we raised many questions.  What accounts for the difference
	between theory and experiment, and between different experiments?  Is the theory
	incomplete or inaccurate at this
	level of description? Experimentally, we do not yet know what material 
	features are required to produce
	universality of the pulse shape function. Differences between experimental results 
	make further experimental tests desirable (see Fig. 5).  

	Although the three experiments we 
	examined do not fall into the same universality class and do not have
	very similar pulse shapes, they share one
	universal aspect not shared by the results from simulation: all the
	experimental averaged pulse shapes are asymmetric in the leftward
	direction.  This suggests that there may be a phenomena that exists
	in real materials that has not been accounted for in theory.

\begin{center}\bf ACKNOWLEDGMENTS \end{center}

	We would like to thank M. B. Weissman, G. Durin, A. Travesset, and 
	R. A. White for very useful discussions. We also thank M. Kuntz and J. Carpenter 
	for providing the front propagation model simulation code.
	K.D. and A.P.M. acknowledge support from NSF via Grant Nos. DMR 99-76550,
	the Materials Computation Center, through NSF Grant No. 99-72783, and IBM 
	which provided the computers that made the simulation work
	possible. J.P.S. acknowledges support from NSF via Grant No. DMR 98-73214.
	A.C.M. acknowledges support from NSF via Grant No. DMR 99-81869.
	A.P.M. would also like to acknowledge the support provided by UIUC through a
	University Fellowship, and K.D. gratefully acknowledges support through an
	A.P. Sloan fellowship.

\begin{figure}
\centerline{
\epsfxsize=3.3in
\epsfbox{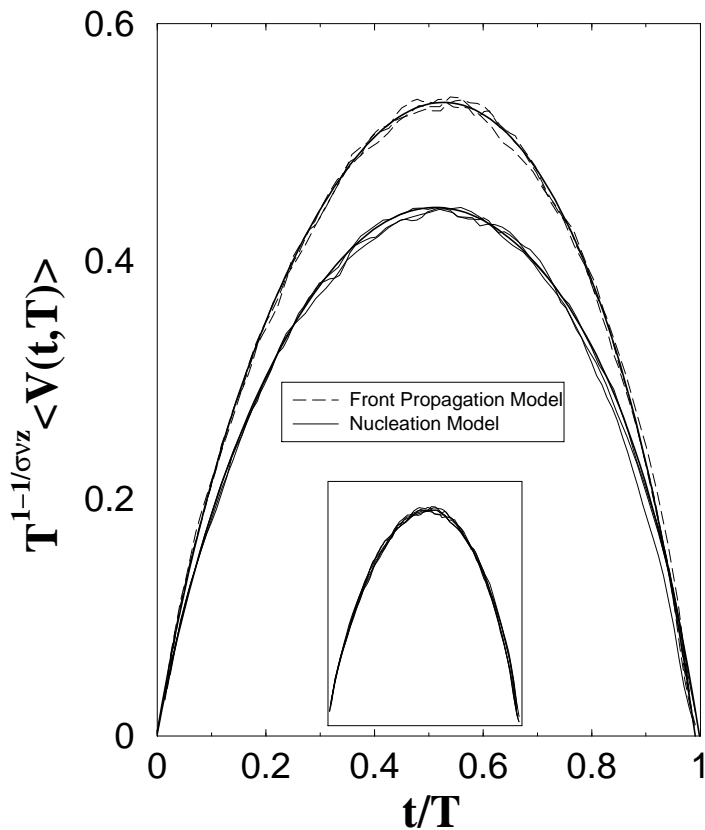}}
\vspace*{0.2in}
\caption{Pulse shape collapses of $V(t,T)$ obtained from simulation. Three
	pulse shapes were collapsed for each model; these pulse shapes represent 
	averaged avalanches of pulse durations 
	$T = 52$, $73$, and $106$ within $5\%$.  From the collapse of front propagation model 
	avalanche pulse shapes, we find $1/\sigma\nu z = 1.72 \pm 0.03$. From the collapse of nucleation
	model avalanche pulse shapes, we find $1/\sigma\nu z = 1.75 \pm 0.03$. The bold line
	going through the collapses is the non-linear curve fit obtained from the set
	of orthonormal polynomials presented in this paper (Eqns.(2)-(6)). Note that the non-linear curve
	fit is shown for only one of the collapsed averaged avalanches in each case. Inset: By rescaling 
	the height of the nucleation model collapse by $20\%$ we obtain a collapse of pulse shapes from
	the two different models suggesting that their pulse shapes are very similar, but quantitatively not
	the same as described in the text. }
\label{fig:1}
\end{figure}

\begin{figure}
\centerline{
\epsfxsize=3.3in
\epsfbox{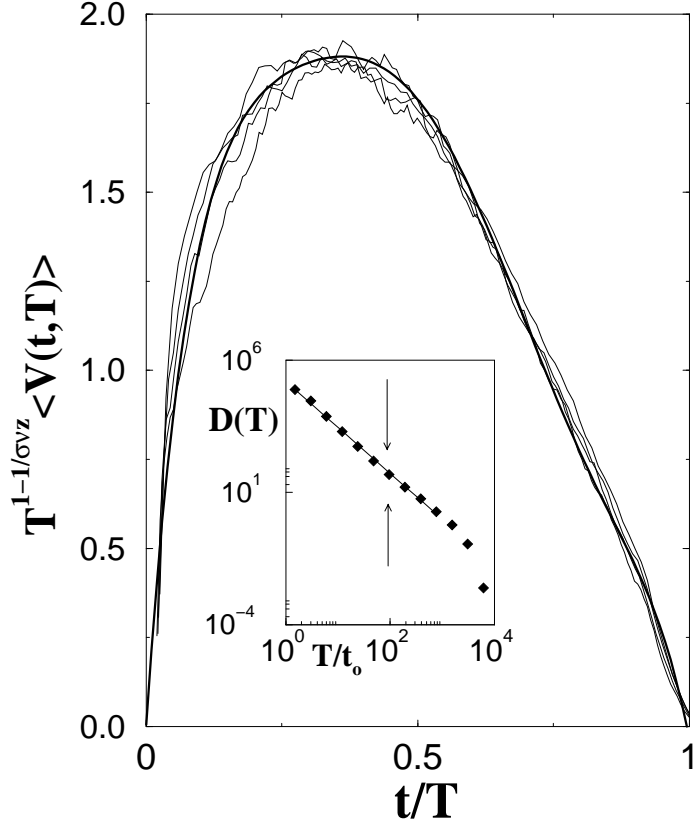}}
\vspace*{0.2in}
\caption{Collapse of averaged experimental avalanche pulse shapes is shown, yielding 
	$1/\sigma\nu z = 1.70 \pm 0.05$.  The four curves represent averaged avalanches 
	of pulse duration: $T =  88t_o$, $110t_o$, $132t_o$,
	and $165t_o$ within $10\%$, where $t_o = 6.4\mu s$ represents the time between 
	each measurement of the Barkhausen noise train. Each of the four curves is
	an average of between 1152 to 1561 avalanches.
	The smooth bold curve is a fit of the averaged avalanche of duration $T = 132t_o$ 
	using the orthonormal polynomials given in Eqns.(2)-(6). 
	Inset: Distribution of avalanche pulse durations,
	$D(T)$, obtained from our experiment. Avalanche pulse shapes were extracted from the
	region indicated by the arrows, and this region is well within the scaling regime. In this 
	scaling regime $D(T)$ scales as $D(T) \sim T^{-\alpha}$ where $\alpha = 1.74\pm0.06$.
	This value of $\alpha$ is between the mean field value of $\alpha = 2.0$ 
	(sample without stress) and the value of $\alpha = 1.5$ for front propagation
	dynamics (sample with stress)\protect\cite{Durin2}, indicating that our experiment may be in a 
	crossover regime.}
\end{figure}

\begin{figure}
\centerline{
\epsfxsize=3.3in
\epsfbox{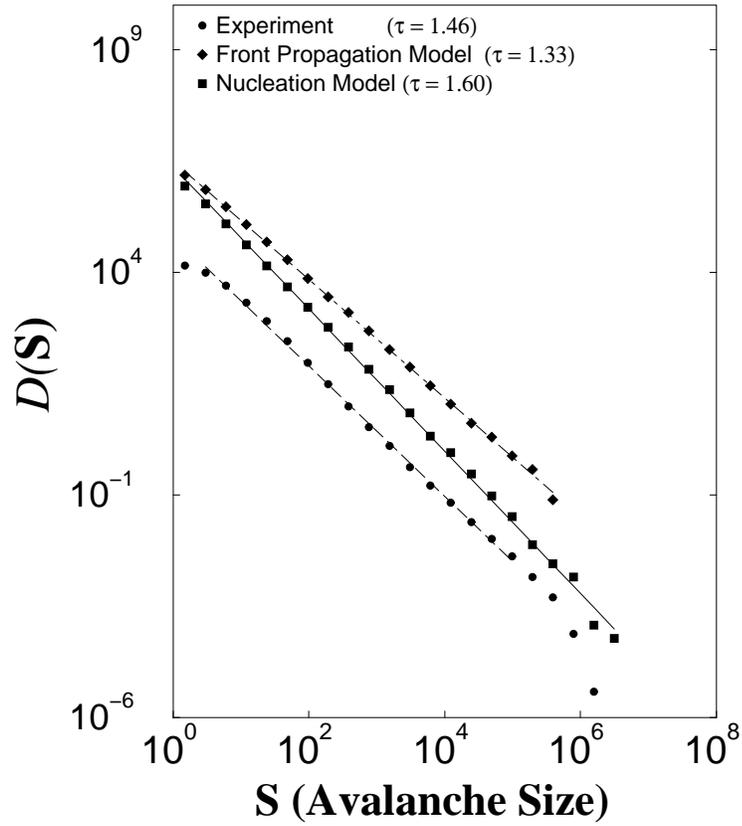}}
\vspace*{0.2in}
\caption{ Avalanche size distribution for front propagation model, nucleation model, and our experiment.
	  The exponent $\tau$ given in the legend is the critical exponent corresponding to the scaling
	  of the avalanche size distribution $(D(S) \sim S^{-\tau})$. }
\label{fig:3}
\end{figure}

\begin{figure}
\centerline{
\epsfxsize=3.3in
\epsfbox{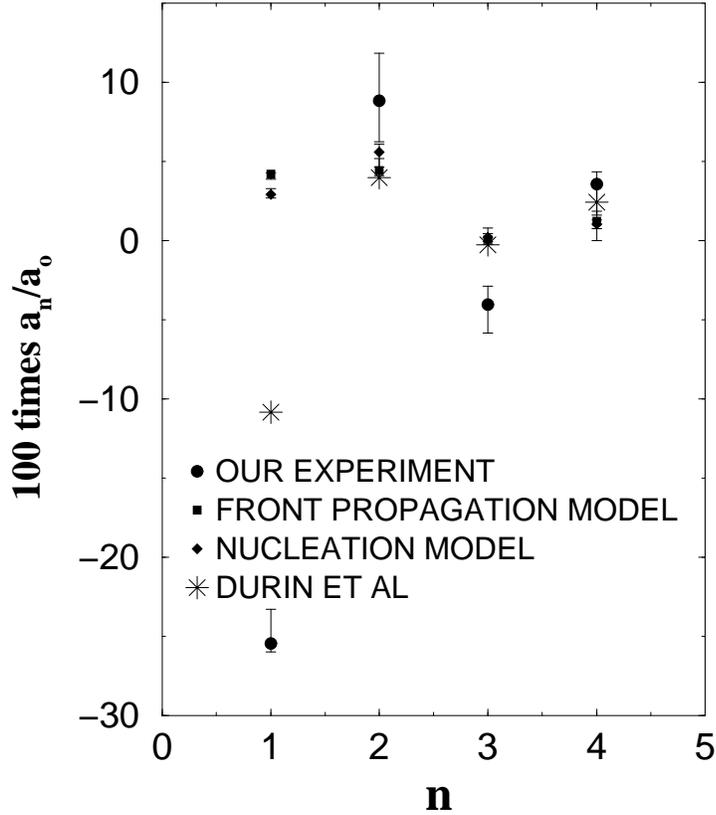}}
\vspace*{0.2in}
\caption{ Fitting coefficients to the avalanche shapes determined for the two models, our
	  experiment, and Durin et al's experiment\protect\cite{Durin}. We find that the 
	coefficients are very similar for the two models.
	While the $a_1$ coefficients determined from the experiments are significantly different from
	the two models, the difference is not only in the sign of asymmetry but also in
	the magnitude of asymmetry. Each fitting coefficient, except for Durin et al's, was 
	determined from three realizations of the universal scaling function in each case.  
	The coefficients, for the two models and our experiment, plotted above, represent 
	median values, while the error bars are determined from the higher and lower values. 
	Durin et al's avalanche shape, presented in Fig. 5, was used to calculate the coefficients
	presented above; no error bars are provided in this case since only one realization was
	available. }
\label{fig:4}
\end{figure}

\begin{figure}
\centerline{
\epsfxsize=3.3in
\epsfbox{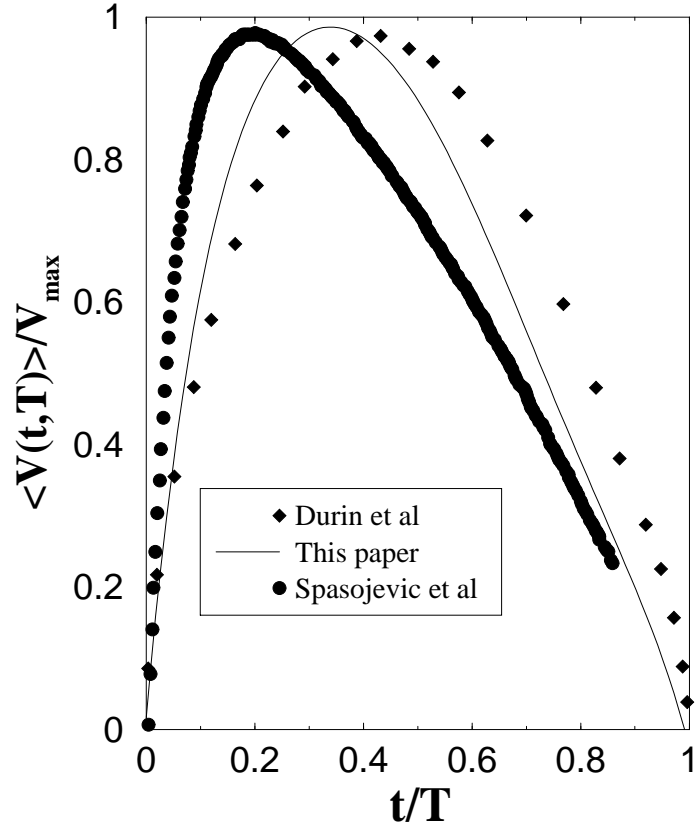}}
\vspace*{0.2in}
\caption{ Comparison of our experimental pulse shapes with experimental pulse shapes obtained by
	  two other groups. Durin et al's sample was under stress with large stress
	  induced anisotropies, putting their experiment into the universality class of front 
	  propagation\protect\cite{Durin2,Durin3,Robbins}. For Durin's experiment $1/\sigma\nu z = 1.77$
	  \protect\cite{Durin} and $\tau = 1.27$\protect\cite{Durin2,Durin3}.
	  For Spasojevi\'{c}'s experiment $1/\sigma\nu z = 1.58$ and $\tau = 1.77$
	  \protect\cite{Spaso}. }
\label{fig:5}
\end{figure}

\renewcommand{\arraystretch}{0.75}
\begin{table}
\begin{tabular}{|c||c|c|c|c|}
&  &  & Our Experiment &  \\ 
& Nucleation & Front Propagation & Durin et al's Expt.$^{\dag}$ 
& Mean Field \protect\cite{Matt2} \\
& Model & Model & Spasojevi\'{c} et al's Expt.$^*$ & \\
	\hline
               & $1.60 \pm 0.04$ & $1.33 \pm 0.08$ & $1.46 \pm 0.05$ & \\
        $\tau$ & $1.60 \pm 0.06$\protect\cite{Olga1} & 1.28\protect\cite{Matt2} 
	       & $1.27 \pm 0.03^\dag$\protect\cite{Durin2,Durin3} & $1.5$ \\
	       & & & $1.77^*$\protect\cite{Spaso} & \\
	\hline
	       & $1.75 \pm 0.03$ & $1.72 \pm 0.03$ & $1.70\pm0.05$ & \\
	$1/\sigma\nu z$ & $1.75 \pm 0.07$\protect\cite{Olga1} & 1.72\protect\cite{Matt2} 
	       & $1.77\pm0.12^\dag$ \protect\cite{Durin2,Durin} & $2$  \\ 
	       & & & $1.58^*$\protect\cite{Spaso} & \\
\end{tabular}
\vspace*{0.2in}
\caption{ In the above table we present the critical exponents 
	  $\tau$ and $1/\sigma\nu z$ in $d = 3$ dimensions for the nucleation model,
	  front propagation model, and all three experiments discussed in this paper.  
	  We also include the mean field values of these exponents. } 
\end{table}

\end{document}